\documentclass[12pt]{iopart}
\usepackage{graphicx,hyperref}
\usepackage{iopams}
\usepackage[small]{caption}

\begin{document}

\title{Mass effect and coherence in medium-induced QCD  radiation off a $q\bar q$ antenna}

\author{N\'estor Armesto, \underline{Hao Ma}, Yacine Mehtar-Tani, Carlos A. Salgado}

\address{Departamento de F\'isica de Part\'iculas and IGFAE, Universidade de Santiago de Compostela, E-15782 Santiago de Compostela, Galicia-Spain}

\author{Konrad Tywoniuk}

\address{Department of Astronomy and Theoretical Physics, Lund University, S\"olvegatan 14A, S-223 62 Lund, Sweden}

\begin{abstract}
The medium-induced one-gluon radiation spectrum off a massive quark-antiquark ($q {\bar q}$) antenna traversing a colored QCD medium is calculated in this contribution. 
The gluon spectrum off the antenna computed at first order in the opacity expansion is collinear finite but infrared divergent, which is different from the result obtained from an independent emitter which is both infrared and collinear finite. The interference between emitters dominates the soft gluon radiation when the antenna opening angle is small and the emitted gluon is soft, whereas the antenna behaves like a superposition of independent emitters when the opening angle is large and the radiated gluon is hard.
As a phenomenological consequence, we investigate the energy lost by the projectiles due to the radiation.
In general, the size of the mass effects is similar in both cases.
\end{abstract}


\section{Introduction}
Calculations of radiative energy loss via medium-induced gluon radiation (see e.g. \cite{CS} and refs. therein) off massive quarks were performed in \cite{ASW, DGLV,WW}. Two effects compete against each other. The gluon formation time, $\left[t_g^{\rm form}(m\ne 0)\right]^{-1}\sim \left[t_g^{\rm form}(m= 0)\right]^{-1}+m_q^2/E_q$, of a massive quark (with mass $m_q$ and energy $E_q$) is shorter than that of a massless quark, so gluon emission in the former is less suppressed by the LPM effect than that in the latter. On the other hand, gluon radiation off a massive quark is suppressed at angles smaller than the so-called dead cone angle $\theta_0$ $=$ $m_q / E_q$, similarly to the case in vacuum \cite{dc,DeadCone}.  The generic result of this competition is
 that a massive quark loses less energy  in a medium than a massless one.
But the mentioned calculations do not consider the effects of the interference between emitters and therefore the extension to multi-gluon radiation relies on ad hoc conjectures.

In order to investigate these interference effects in vacuum, a quark-antiquark ($q {\bar q}$) antenna was considered, see \cite{book} and references therein. The soft gluon radiation spectrum off a massless $q {\bar q}$ antenna in vacuum exhibits angular ordering (AO): radiation is suppressed at $\theta$ $>$ $\theta_{q {\bar q}}$ (after averaging over the gluon azimuthal angle), where $\theta$ is the angle between the emitted gluon and the parent quark (antiquark) and $\theta_{q {\bar q}}$ is the antenna opening angle. The massless antenna spectrum diverges when $\theta$ $\rightarrow$ $0$ and $\omega$ $\rightarrow$ $0$.
When the quarks have a non-zero mass,  AO is modified and the collinear divergence disappears due to the dead cone effect, but the soft divergence remains. The result was shown to hold for arbitrary color representations of the $q\bar q$ pair.

Medium-induced soft gluon radiation off a massless $q {\bar q}$ antenna has been studied recently \cite{MSTprl, MSTdecoh, MT, CI}. The spectrum exhibits antiangular ordering (AAO) - there is no collinear divergence ($\theta$ $>$ $\theta_{q {\bar q}}$ $>$ $0$) - but the soft divergence persists, at variance with the results in \cite{ASW, DGLV,WW}.
Due to space limitations, here we show some selected results for a massive quark-antiquark pair at first order in opacity.

\section{Results}
\subsection{Considered diagrams}

\begin{figure}[ht]
\begin{center}
\includegraphics[width=0.8\textwidth]{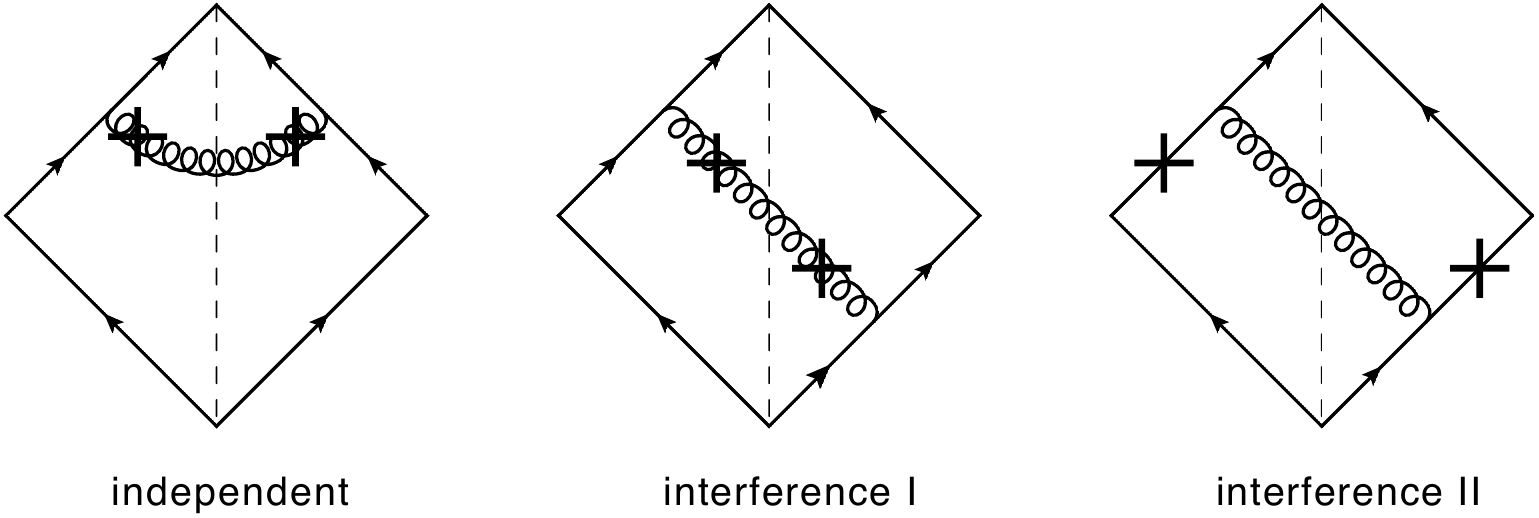}
\caption{\label{aspec} Examples of Feynman diagrams representing the three contributions to the antenna spectrum. The cross denotes the position of scattering, the l.h.s. of the dashed line is the amplitude and the r.h.s. of the dashed line is the complex conjugate amplitude, with the dashed line being the cut.}
\end{center}
\end{figure}

In Fig. \ref{aspec} we show the three types of diagrams that we consider. They correspond to the independent emission off the quark (antiquark), denoted as  independent and the interference between the quark and antiquark, denoted as interference I and interference II, respectively. The total number of diagrams computed is 96.

%

\subsection{Average energy loss}

The average radiative energy loss for gluons with energies $\omega_{\rm min} < \omega <\omega_{\rm max}$ is 
\begin{center}
\begin{equation}
\Delta E = \int_{\omega_{\rm min}}^{\omega_{\rm max}} {\rm d} \omega \int_0^{\pi / 2} {\rm d} \theta \, \omega \frac{{\rm d} N}{{\rm d} \omega \, {\rm d} \theta}\ .
\label{deoe}
\end{equation}
\end{center}

\begin{figure}[h]
\begin{center}
\includegraphics[width=\textwidth]{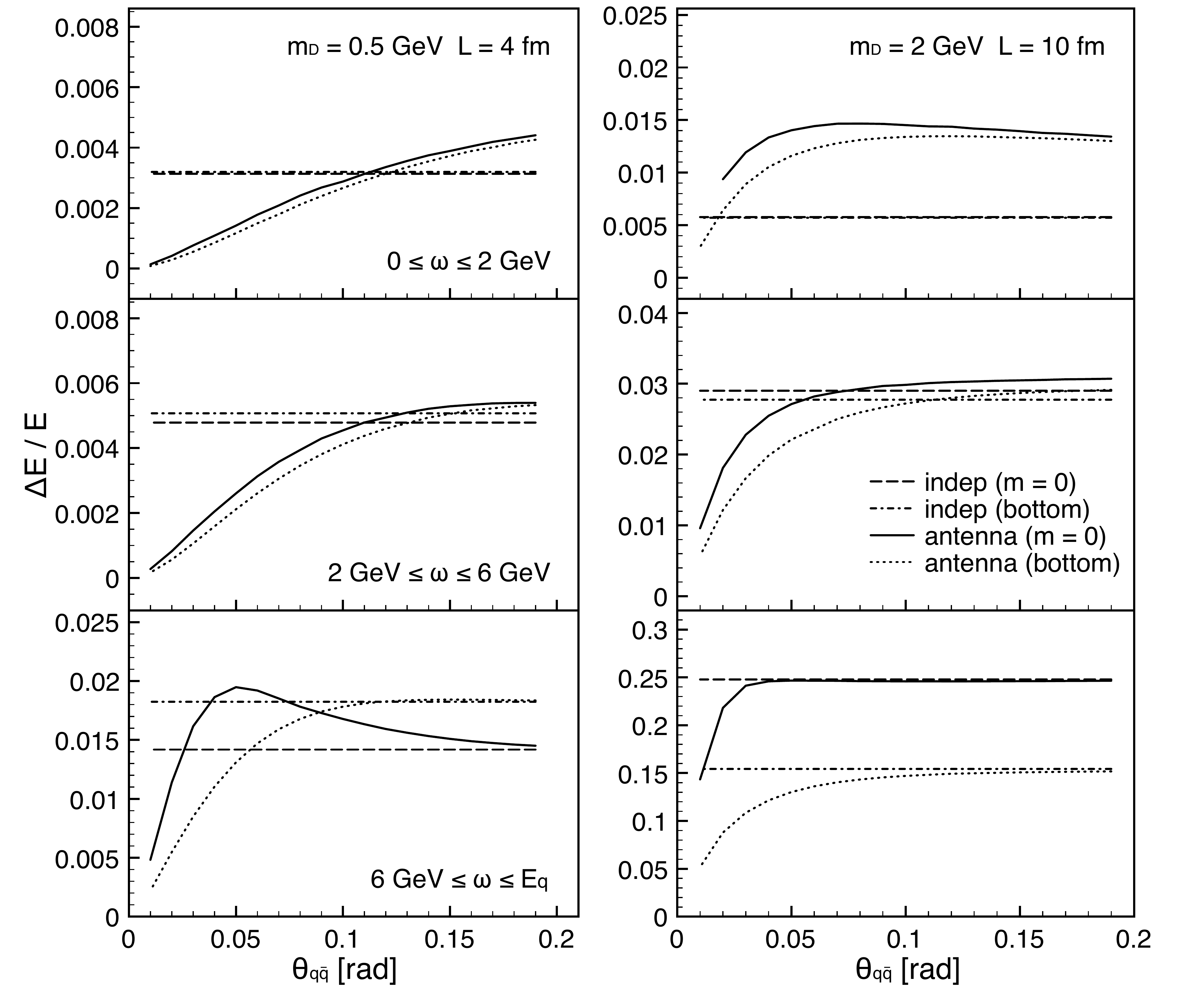}
\caption{\label{HOA}Dependence of the medium-induced radiative relative energy loss on the antenna opening angle. The parameters are: Debye mass $m_D=0.5(2)$ GeV and medium length $L=4(10)$ fm for the plots on the left (right). The solid curves correspond to the massless antenna, the dotted curves to the bottom antenna, the dashed curves to the massless independent spectra and the dash-dotted curves to the bottom independent spectra. From top to bottom, the values used in Eq. (\ref{deoe}) for $\omega_{\rm min}$ are 0, 2 and 6 GeV, while those for $\omega_{\rm max}$ are 2, 6 and $E_q$.}
\end{center}
\end{figure}

The ratio $\Delta E / E$ as a function of $\theta_{q {\bar q}}$ is shown in Fig. \ref{HOA}, where  $E=E_q=100$ GeV for the independent emitter case and for the antenna the spectrum is divided by 2 as it gets contributions from both the quark and antiquark. 
In the soft and moderate gluon energy regions ($0\leq\omega\leq 2$ GeV and $2$ GeV $\leq \omega\leq 6$ GeV), both the massless and the massive antenna spectra grow monotonously with an increasing opening angle $\theta_{q {\bar q}}$ and the former is larger than the latter. In the hard gluon radiation sector ($6$ GeV $\leq\omega\leq E_q$), the situation is similar for large medium parameters while, for small medium parameters (lower left plot), there is a crossing between the massless and the bottom antennas due to the LPM effect which results in a larger energy loss for larger masses as discussed in the Introduction and found previously e.g. in \cite{ASW}. Therefore, for large medium parameters the dead cone effect dominates over the LPM in all cases.
Both massless and massive antenna results approach the ones from independent emitters when $\theta_{q {\bar q}}$ is large, showing that the interference between the quark and the antiquark of the antenna reduces with an increasing opening angle. 
Besides, in the soft gluon emission region and for large medium parameters, there is apparently more energy loss in the antenna than for independent emitters in both the massless and the massive cases. This reflects the fact that the antenna spectrum exhibits a soft divergence while the independent spectrum is infrared finite. In the moderate and the hard gluon emission sectors, the antenna average energy loss increases and gradually approaches the independent average energy loss with an increasing antenna opening angle $\theta_{q {\bar q}}$, which indicates that more collimated projectiles lose less energy. There is no radiation for the antenna when $\theta_{q {\bar q}}$ $\rightarrow$ $0$. The size of the mass effect in the antenna is similar to the one for independent emitters.

\section{Conclusions}

In this contribution, we show some results of the medium-induced gluon radiation spectrum off a $q {\bar q}$ antenna at first order in opacity for the massive case. In this computation, performed in the high-energy limit, both the non-abelian LPM effect and the dead cone for massive quarks (both contained in the medium-induced gluon spectrum off individual emitting partons - the BDMPS-Z-W/GLV formalism), and the interference between emissions off the quark and antiquark, are included.

The antenna radiation is found to be dominated by that off independent emitters for large opening angles of the antenna and for large energies of the emitted gluon. More collimated antennas lose less energy. The phase space restriction for gluon emission implied by the dead cone effect  is similar in the antenna and in the case of independent emitters. The effect of the interference between different emitters is to generate predominantly soft radiation at large angles.

\vskip 0.3 cm

\small{The work of NA, HM, YM-T, and CAS is supported by Ministerio de Ciencia e Innovaci\'on
of Spain (grants FPA2008-01177 and FPA2009-06867-E), Xunta de Galicia (Conseller\'{\i}a
de Educaci\'on and grant PGIDIT10PXIB 206017PR),  project Consolider-Ingenio 
CPAN CSD2007-00042, and FEDER. The work of KT is supported  by the Swedish Research Council (contract number 621-2010-3326). CAS is a Ram\'on y Cajal researcher.}

\end{document}